\documentclass[useAMS, usenatbib]{mn2e} 
\usepackage{graphicx}

\title{Mass Accretion Rates and Histories of Dark Matter Haloes }

\author[J McBride, O Fakhouri and C-P Ma]{James McBride\thanks{E-mail: jmcbride@berkeley.edu, onsi@berkeley.edu, cpma@berkeley.edu}, Onsi Fakhouri, and Chung-Pei Ma\\
Department of Astronomy, 601 Campbell Hall, University of California, Berkeley, CA 94720}

\pagerange{ 
\pageref{firstpage}-- 
\pageref{lastpage}}

\usepackage{amsmath} 
\usepackage{multirow} 
\usepackage{float} 
\usepackage{array}

\newcommand{\ximin}{\xi_{\rm min}} 
\newcommand{\dRfof}{\delta_{R-{\rm FOF}}}

\begin{document}

\label{firstpage}

\maketitle 
\begin{abstract}
  We use the extensive catalog of dark matter haloes from the Millennium
  simulation to investigate the statistics of the mass accretion histories
  (MAHs) and accretion rates of $\sim 500,000$ haloes from redshift $z=0$
  to 6. We find only about 25\% of the haloes to have MAHs that are well
  described by a 1-parameter exponential form. For the rest of the haloes,
  between 20\% (Milky-Way mass) to 50\% (cluster mass) experience late-time
  growth that is steeper than an exponential, whereas the remaining haloes
  show plateau-ed late-time growth that is shallower than an
  exponential. The haloes with slower late-time growth tend to reside in
  denser environments, suggesting that either tidal stripping or the
  ``hotter'' dynamics are suppressing the accretion rate of dark matter
  onto these haloes. These deviations from exponential growth are well fit
  by introducing a second parameter: $M(z) \propto (1+z)^\beta e^{-\gamma
    z}$. The full distribution of $\beta$ and $\gamma$ as a function of
  halo mass is provided. From the analytic form of $M(z)$, we obtain a
  simple formula for the mean accretion rate of dark matter, $\dot{M}$, as
  a function of redshift and mass. At $z=0$, this rate is $42 M_\odot\,
  {\rm yr}^{-1}$ for $10^{12} M_\odot$ haloes, which corresponds to a mean
  baryon accretion rate of $\dot{M}_b=7 M_\odot \, {\rm yr}^{-1}$. This
  mean rate increases approximately as $(1+z)^{1.5}$ at low $z$ and
  $(1+z)^{2.5}$ at high $z$, reaching $\dot{M}_b=27, 69$, and 140
  $M_\odot\, {\rm yr}^{-1}$ at $z=1$, 2, and 3. The specific rate depends
  on halo mass weakly: $\dot{M}/M \propto M^{0.127}$. Results for the broad
  distributions about the mean rates are also discussed.
\end{abstract}

\section{Introduction} \label{introduction}

The mass growth history is a basic property of dark matter haloes. Haloes
in numerical simulations are seen to be assembled through a number of
processes: mergers with comparable mass haloes (``major mergers''), mergers
with smaller satellite haloes (``minor mergers''), and accretion of
non-halo material that is composed of either haloes below the numerical
resolution or diffuse particles. Following the mass history of the most
massive progenitor halo as a function of redshift $z$ is a useful way to
quantify a halo's mass assembly history. These mass accretion histories
(MAHs, or $M(z)$) are important for statistical studies of the
distributions of halo formation redshifts, and the correlations between
formation time and other halo properties such as environment,
concentration, substructure fraction, spin, and relative contributions to
mass growth from major vs minor mergers. Moreover, the time derivative of
the MAH gives the mass growth rate of dark matter haloes, which is directly
related to the accretion rate of baryons from the cosmic web onto dark
matter haloes.

A number of earlier papers have investigated various aspects of the halo
MAHs. For instance, \citet{Wechsler02} analyzed $\sim 900$ haloes (above
$10^{12} h^{-1} M_\odot$ at $z=0$) in a $\Lambda$CDM simulation in a
$60\,h^{-1}$ Mpc box with $256^3$ particles. The values from a 1-parameter
fitting function for the MAHs were presented for 8 haloes. Clear
correlations between the formation redshift $z_f$ and concentration $c$ of
haloes were seen, with late-forming haloes being less concentrated. The
scatter in $c$ was attributed to the scatter in $z_f$. An alternative
2-parameter fitting function was demonstrated by \citet{VDBosch02} to be
superior to a 1-parameter fit to haloes in a simulation with the same
particle number in a $141 h^{-1}$ Mpc box.

The relationship between halo structure and accretion was further addressed
in \citet{Zhao03mnras} and \citet{Zhao03apj}, where the redshift dependence
of $c$ was observed to be more complicated than a simple
proportionality. \citet{Tasitsiomi04} examined 14 haloes, ranging in mass
from group to cluster scale (.58 to $2.5\times 10^{14} h^{-1} M_\odot$) and
also found that a 2-parameter fit for $M(z)$ worked better. \citet{Cohn05}
studied the mass accretion histories of $\sim 1500$ cluster-sized haloes
and characterized several properties of galaxy cluster formation.

\citet{Maulbetsch07} studied the environmental dependence of the formation
of $\sim 4700$ galaxy-sized haloes (above $10^{11} h^{-1} M_\odot$) in a
$50\, h^{-1}$ Mpc simulation box. In higher-density environments, they
found the haloes to form earlier with a higher fraction of their final mass
gained via major mergers. \citet{limogao08} studied 8 different definitions
of halo formation time using the haloes from the Millennium simulation
\citep{Springel05}. The motivation was to search for halo formation
definitions that better characterize the downsizing trend in star formation
histories, as opposed to the hierarchical growth of haloes in the
$\Lambda$CDM cosmology. \citet{Zhao08} (Z08) investigated the mean MAH in
different cosmological models -- scale-free, $\Lambda$CDM, standard CDM,
and open CDM -- and searched for scaled mass and redshift variables that
would lead to a universal fitting form for the median MAH for all models.

The results in these earlier papers were presented either for $M(z)$ of a
handful of individual haloes, or for the global mean growth of a selection
of haloes. Our aim here is to quantify systematically the diversity of
growth histories and rates using the $\sim 500,000$ $z=0$ haloes with
$M>10^{12} M_\odot$ (i.e. above 1000 simulation particles) and their
progenitors in the Millennium simulation. Over this large range of haloes,
we find that an exponential fit does not adequately capture the behavior of
halo growth. Many haloes experience large changes in the rate at which they
accrete mass. Some haloes grow more slowly at late times, and occasionally
even lose mass, while other haloes undergo late bursts of growth. All of
these MAHs are poorly fit by an exponential, and suggest the need for a
fitting form with more flexibility. We find it helpful to classify the MAHs
into four types based on their late-time accretion rate. The large ensemble
of haloes allows us to quantify the mean values as well as the dispersions
of the mass accretion rates and halo formation redshifts as a function of
mass and redshift.

This paper is organized as follows. Sec.~2 provides some background
information about the haloes in the Millennium simulation and describes how
we construct halo merger trees. This post-processing of the Millennium
public data is necessary for identifying the thickest branch (i.e. the most
massive progenitor) along each final halo's past history. The masses of
these progenitors will then allow us to quantify the MAH, $M(z)$. In
Sec.~3, we first assess the accuracy of the 1-parameter exponential form
for $M(z)$. We then propose a more accurate two-parameter function for
$M(z)$ and classify the diverse assembly histories into four broad types
according to their late-time growth behavior.  We further quantify the
statistics of the two fitting parameters, providing (in the Appendix)
algebraic fits for their joint distributions that can be used to generate
Monte Carlo realizations of an ensemble of halo growth tracks. The
applicability of $M(z)$, which is derived for $z=0$ haloes, for the mass
accretion history of higher-redshift haloes is discussed in
Sec.~3.3. Sec.~4 is focused on the statistics of the mass accretion
rates. A simple analytic expression is obtained for the mean accretion
rate, $\left< \dot{M} \right>$, of dark matter as a function of halo mass
and redshift. The dispersions about the mean rates are significant, as
evidenced by the differential and cumulative distributions of $\dot{M}$
presented here. Sec.~5 discusses the mean and the distribution of the halo
formation redshift as a function of halo mass. In Sec.~6 we report the
correlations of MAHs with halo environment, the last major merger redshift,
and the fraction of haloes' final masses assembled via different types of
mergers.

\section{Halo Merger Trees in the Millennium Simulation} \label{background}

The Millennium simulation \citep{Springel05} provides a database for the
evolution of roughly $2\times 10^7$ $z=0$ dark matter haloes from redshifts
as high as $z=127$ in a $500 h^{-1}$ Mpc box using $2160^3$ particles of
mass $1.2\times 10^9 M_\odot$ (all masses quoted in this paper are in units
of $M_\odot$ and not $h^{-1} M_\odot$). It assumes a $\Lambda$CDM model
with $\Omega_m=0.25$, $\Omega_b=0.045$, $\Omega_\Lambda=0.75$, $h=0.73$,
and a spectral index of $n=1$ for the density perturbation power spectrum
with a normalization of $\sigma_8=0.9$.

Dark matter haloes are identified with a friends-of-friends (FOF) group
finder \citep{Davis85} with a linking length of $b=0.2$. Throughout this
paper we use the number of particles linked by the FOF finder to define the
halo's mass. Once identified, each FOF halo is then broken into
gravitationally bound substructures (subhaloes) by the SUBFIND algorithm
(see \citealt{Springel01SUBFIND}). These subhaloes are connected across the
64 available redshift outputs: a subhalo is the descendant of a subhalo at
the preceding output if it hosts the largest number of the progenitor's
bound particles. The resulting subhalo merger tree can be used to construct
merger trees of FOF haloes, although we have discussed at length in
\citet{FM08, FM09} the complications due to halo fragmentation
and have presented comparisons of several post-processing algorithms that
handle fragmentation events.

Our results in this paper are based on the stitch-3 post-processing
algorithm described in \citet{FM08}. In this algorithm, fragmented haloes
that remerge within 3 outputs after fragmentation are stitched into a
single FOF descendant; those that do not remerge within 3 outputs are
snipped and become orphan haloes. After applying the stitch-3 algorithm, we
extract the mass accretion history, $M(z)$, of each halo at $z=0$ (or at
any higher redshift) by following the halo's main branch of progenitors. We
have compared the resulting $M(z)$ and formation redshifts to those
obtained from the alternative algorithms (e.g., ``snip,'' ``split,'' and
subhalo vs FOF mass) discussed in \citet{FM08, FM09}.  We find the
systematic variations to all be within 5-10\% of the stitch-3 values of
these quantities.

\section{Fitting Mass Accretion Histories} \label{MAH_fit}

\subsection{Previous MAH Forms} \label{old_fits}

To quantify the limitations of the exponential fit in capturing halo
growth, consider the formation redshift $z_f$, here defined as the redshift
at which $M(z)$ is equal to $M_0/2$.

For the 1-parameter exponential form (e.g. \citealt{Wechsler02}) 
\begin{equation}
	M(z) = M_0e^{-\alpha z}, 
\end{equation}
the parameter $\alpha$ is simply related to $z_f$ by 
\begin{equation}
	z_f = \frac{\ln(2)}{\alpha}. 
\end{equation}
We have compared $z_f$ as determined by the exponential fit to each halo's
$M(z)$ from the simulation with the $z_f$ determined directly from the
$M(z)$ tracks such that $M(z_f)=M(0)/2$ (using interpolation between output
redshifts).
We find the exponential fit to err systematically in its determination of
$z_f$, significantly overestimating the formation redshift for haloes that
form recently and underestimating it for haloes that form early.  The mean
value of $z_f$ from the exponential fit, for instance, is 0.3 higher than
the actual value for young haloes and is 0.8 lower for old haloes across
all masses.


A more complicated functional representation of MAHs was put forth by
\citet{VDBosch02}:
\begin{equation}
  \log\left(\frac{M(z)}{M_0}\right) = - 0.301 
    \left[\frac{\log(1 + z)}{\log(1 + z_f)}\right]^\nu, 
\end{equation}
where $\nu$ and $z_f$ are fitting parameters. The use of an additional
parameter provided significant improvement in the quality of the fits for
many MAHs, especially those that formed late.
This 2-parameter form, however, is not flexible enough to handle haloes
that have lost mass, as it cannot take on values that would give $M(z)/M_0$
greater than 1. Moreover, over the sample of haloes tested in
\cite{VDBosch02}, comparison between the goodness-of-fit of this
two-parameter form and the exponential fit showed that the exponential fit
actually performs better for early forming haloes.

\subsection{A Revised MAH Form} \label{revised_fit}

To address the need for a fit that is both effective and simple, we find a
2-parameter function of the form
\begin{equation}
	M(z) = M_0 (1 + z)^\beta e^{-\gamma z}, \label{ourfit} 
\end{equation}
to be versatile enough to accurately capture the main features of most MAHs
in the Millennium Simulation.  
This form has also been studied in \citet{Tasitsiomi04} for cluster-mass
haloes, but it has not been tested over a large number of haloes of
different mass.  
The form reduces to an exponential when $\beta$ is 0, and $\gamma$ in this
case is simply the inverse of the formation redshift: $\gamma=
\ln(2)/z_f$. A large fraction of the haloes, however, are better fit when
the additional factor of $(1+z)^\beta$ in equation~(\ref{ourfit}) is
included.  In general, $\beta$ can be either positive or negative, but $\gamma \ge 0$.
We find the combination $\beta - \gamma$ to be a useful parameter for
characterizing these MAHs as $\beta-\gamma$ gives the mass growth rate at
small redshifts:
\begin{equation}
  \frac{d\ln(M(z))}{dz} = \frac{\beta}{1 + z} -\gamma \approx \beta -
    \gamma + {\cal O}(z) \,.
\label{eq:mod_deriv} 
\end{equation}
This late-time trend can be used to characterize the MAH as described below.

\begin{figure}
  \centering
  \includegraphics{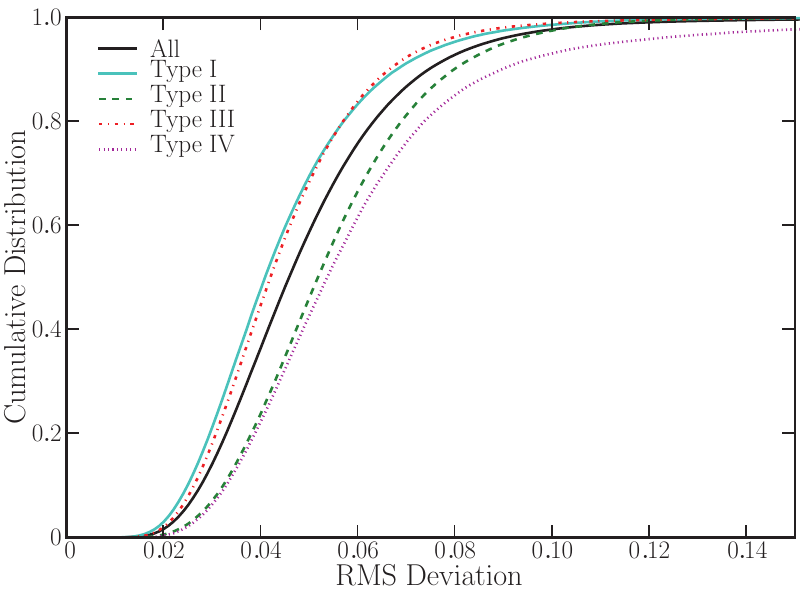}
  \caption{ Cumulative distribution for the RMS deviations between
    the fitting formula eq.~(\ref{ourfit}) and the Millennium Nbody output
    for the mass growths of 478,781 $z=0$ haloes (solid black) and the four
    sub-types (colored curves) listed in Table~1.  The figure shows that
    the fits perform well overall: about 75\% of the haloes have RMS
    deviations less than $\sim 6$\%.} 
\label{fig:rms_dev}
\end{figure}

To obtain the best-fit values for $\beta$ and $\gamma$ in
equation~(\ref{ourfit}), we have performed a $\chi^2$-like minimization of
the quantity
\begin{equation}
    \Delta^2 = \frac{1}{N}\sum_N \frac{ [ M(z_i)/M_0 - 
   (1 + z_i)^\beta e^{-\gamma z_i} ]^2}{M(z_i)/M_0} \,,
\label{Delta}
\end{equation}
where the sum is over the $N$-available simulation redshift outputs at
$z_i (i=1,...,N)$ for each halo.  The choice of the factor $M(z_i)$ in the
denominator is akin to assuming Poissonian errors for halo masses.  We
found this choice to be a suitable middle ground between minimizing simply
the sum of squares and minimizing the fractional deviation (i.e. with a
factor of $M^2(z)$ in the denominator). The former tended to fit the finely
sampled low-$z$ points well at the expense of the sparsely spaced high-$z$
points, whereas the latter tended to do the opposite.
Equation~(\ref{Delta}), on other hand, provides reasonable fits for the
entire history of the halo growth.  Fig.~\ref{fig:rms_dev} shows the
cumulative distribution for the rms deviation of the fits from the
Nbody data (normalized by $M_0$) for all $\sim 500,000$ $z=0$ haloes.
The deviation is less than 6\% for over 75\% of the haloes, and only a few
percent of haloes have deviations larger than 10\%.  Of this most poorly
fit subset of haloes, nearly half underwent mass loss at late times.  As
expected, the fits become progressively worse at higher redshifts; for over
75\% of haloes, the maximum fractional deviation between the fits and Nbody
results occurs above $z=4$.

\begin{table}
	\centering 
	\begin{tabular}
		{c c c c} Type & Criteria & Characteristics & $\frac{\chi^2_1}{\chi^2_2}$\\
		\hline \hline I & $|\beta|$ $<$ 0.35 & Good exponential & $1.09$ \\
		II & $\beta - \gamma <$ -0.45 & Steep late growth & $1.61$\\
		III & -0.45 $< \beta - \gamma <$ 0 & Shallow late growth & $2.15$\\
		IV & $\beta - \gamma >$ 0 & Late plateau/decline & $3.31$\\ 
	\end{tabular}
	\caption{ MAHs are categorized based upon the best fit parameters $\beta$ and $\gamma$ of equation~(\ref{ourfit}). Categorization is done in order by type; thus MAHs that satisfy the criteria for both Type I and Type II belong to Type I. The right-most column is the mean of $\chi^2_1/\chi^2_2$, the ratio of the $\chi^2$ computed for the 1-parameter exponential form to the $\chi^2$ computed for the 2-parameter form in equation~(\ref{ourfit}).  Values $>1$ imply that the 2-parameter form provides a more accurate fit than the 1-parameter form.} 
\label{table:types} 
\end{table}

\begin{figure}
\centering
\includegraphics{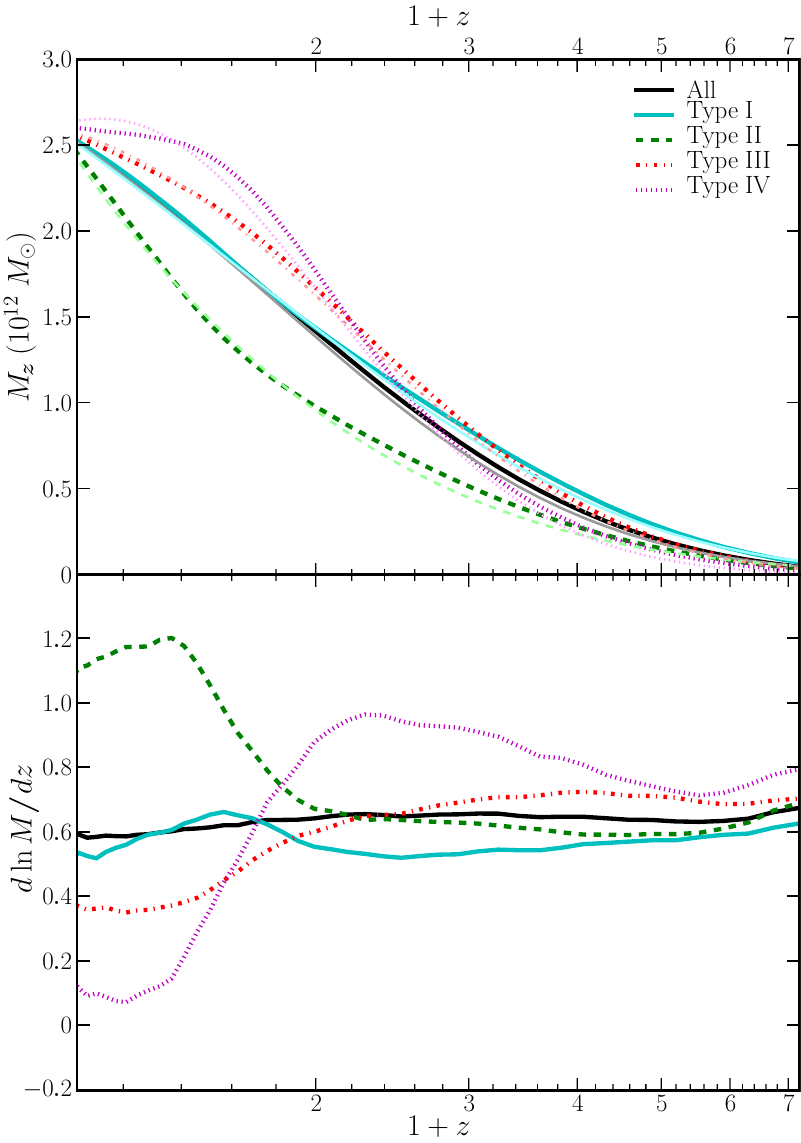}
\caption{Average mass accretion histories (top) and their derivatives
  (bottom) for the total population (black solid) and the four types of
  halo growths listed in Table~\ref{table:types}. The average is taken over
  the $M(z)$ for galaxy-sized haloes with masses between $2.1 \times
  10^{12} M_{\sun}$ and $3.3 \times 10^{12} M_{\sun}$ in the Millennium
  simulation, although haloes at different mass show similar behavior.  In
  the top panel, the set of curves with the lighter shading shows the
  average $M(z)$ computed from our fits of equation~(\ref{ourfit}) to each
  halo's MAH.  The bottom panel illustrates that the late-type growth rates
  differ greatly, ranging from $d\ln M/dz\sim 1.2$ for Type II to 0.1 for
  Type IV. }
\label{fig:mass_sum2}
\end{figure}

We suggest that the parameters $\beta$ and $\gamma$ allow for rough
classifications of MAHs into a few basic groups, summarized in Table
\ref{table:types}. The classification scheme is quite straightforward. Fits
with small values for $\beta$ indicate a weak contribution from the $(1 +
z)$ term, and deviate minimally from an exponential curve. These haloes
with $|\beta| < 0.35$ are labeled Type I.

The rest of the classifications are dependent upon the value of $\beta -
\gamma$. The motivation for this is the fact that the difference represents
the value of the derivative at $z = 0$, as noted in
equation~(\ref{eq:mod_deriv}). Hence Type II haloes, defined to be those
haloes with $\beta - \gamma < -0.45$, feature steep growth at late times,
typically steeper than can be captured by an exponential fit.

Type III haloes have fit parameters that fall in the range $-0.45 < \beta -
\gamma < 0$ and exhibit flat late time growth. Like Type II, these tend to
deviate from the fit that would be found using the exponential form, but
Type III haloes do so in the opposite direction to Type II haloes. A
typical Type III halo has undergone limited growth during recent times,
sometimes after a spurt of growth at earlier times.

Type IV, with $\beta - \gamma > 0$, represents the most extreme deviation
from an exponential. The majority of Type IV haloes have shed mass, some of
them by significant amounts. Some Type IV haloes have merely seen their
growth slow down like the Type III haloes, but over a more significant
period of time. As such, Type IV haloes are extreme cases of Type III
haloes, perhaps representing the future growth for some Type III haloes.

The boundaries delineating these classifications are rough guidelines at
best. For example, consider the definition for Type I of $|\beta| <
0.35$. For the largest values of $\beta$ in this group, which should be
considered the worst of the ``good exponentials'' that constitute Type I,
the fractional difference between the formation redshift as determined by
the simple exponential and the modified exponential is a little under
8\%. The agreement is not perfect, but the two fits are similar enough for
these haloes that the use of the power law parameter adds little. Of
course, there is no reason why we should not instead demand that the
formation redshifts differ on average by no more than 5\%, or perhaps
10\%. In the end, the combination of the formation redshift metric and a
couple of others for comparing the fits suggested that demanding $|\beta| <
0.35$ was inclusive enough to capture the majority of haloes for which an
exponential is an adequate fit, without unduly diminishing the integrity of
the group.

Fig.~\ref{fig:mass_sum2} compares the shapes of the average MAH for haloes
of galaxy-size mass from the Millennium simulation for the overall
distribution and for each type. The bottom panel shows the derivative $d\ln
M/dz$ to highlight the different late-time accretion rates among the four
types. Haloes of other mass show similar behavior. Clearly, the late time
growth rate is an important factor in distinguishing haloes from one
another. The average MAH for Type I haloes is quite similar to that of the
overall distribution, which indicates that the average MAH is approximately
exponential. However, the behavior of about 75\% of individual haloes
deviates from an exponential noticeably.  This fact is quantified in the
right-most column of Table~\ref{table:types}, where the ratio of $\chi^2$
for the exponential fit to the 2-parameter fit is seen to increase with the
MAH types.

Since the mean MAH is approximately exponential, the accretion rate $d\ln
M/dz$ averaged over the whole population is also nearly independent of
redshift (black solid curves in Fig.~\ref{fig:mass_sum2}) when expressed in
units of per redshift, with $d\ln M/dz$ being between 0.6 and 0.7 for $z=0$
up to 5. This weak dependence on redshift is similar to that of the halo
merger rates (per unit $z$) reported in \citet{FM08}. The different types
of haloes, however, show significant dispersions in the late-time accretion
rates, with $d\ln M/dz$ being as high as 1.2 for Type II and as low as 0.1
for Type IV at $z\approx 0$.

For each of the mean profiles shown in Fig.~\ref{fig:mass_sum2}, we have
fit the analytic form in equation~(\ref{ourfit}). The best-fit values of
$(\beta, \gamma)$ are (0.10, 0.69) for all haloes, and ($-0.04$, 0.54),
($-0.9$, 0.35), (0.62, 0.88), and (1.42, 1.39) for each of the four types,
respectively.

\begin{table}
	\centering 
	\begin{tabular}
		{l c c c c c}
		Mass Range & Halo Number & Type I & II & III & IV\\
		($10^{12} M_\odot$) & & & & & \\
		\hline \hline
		1.2 to 2.1 & 191421 & 29\% & 27\% & 32\% & 12\% \\
		2.1 to 4.5 & 143356 & 27\% & 29\% & 32\% & 12\% \\
		4.5 to 14 & 95744 & 24\% & 34\% & 31\% & 11\% \\
		14 to 110 & 43089 & 20\% & 42\% & 26\% & 11\% \\
		$> 110$ & 4787 & 18\% & 57\% & 17\% & 8\% \\
		\hline & 478781 & 27\% & 31\% & 31\% & 11\% \\
	\end{tabular}
	\caption{Within each mass range, the percentage of haloes that belong to each type are provided. For each type, there is a noticeable trend with mass, though the strength of the trend varies.} \label{table:percent_type}
\end{table}

\begin{figure}
  \centering
  \includegraphics{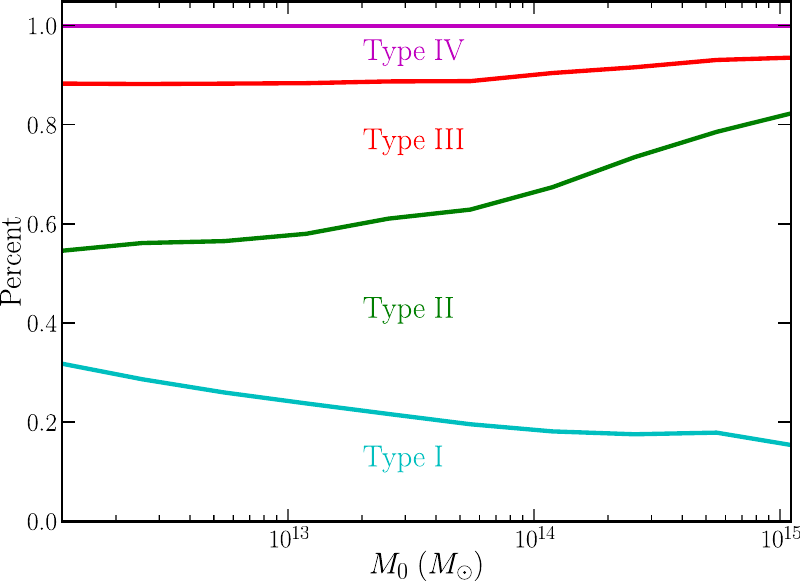}
  \caption{Cumulative fraction of haloes belonging to a given MAH type as a
    function of halo mass (e.g. the magenta type-IV curve includes the
    contributions from the three other types). The exponential form (Type
    I) is a good fit for only 20\% to 30\% of the haloes at all masses. The
    type II fraction shows a strong mass dependence, reaching $\sim 60$\%
    for cluster-mass haloes.}
\label{fig:percent_type}
\end{figure}

The statistics of the 478,781 $z=0$ haloes (above 1000 particles, or a mass
of $1.2\times10^{12} M_\odot$) belonging to each MAH type across different
mass bins are given in Table~\ref{table:percent_type} and
Fig.~\ref{fig:percent_type}. Exponential MAH (Type I) is seen to apply to
only 20 to 30\% of the haloes. There is also interesting dependence of the
type on halo mass. Most notably, Type II haloes feature a strong dependence
on mass, where the fraction rises from 27\% at $\sim 10^{12} M_\odot$ to
60\% at $\ga 10^{14} M_\odot$. Cluster-size haloes therefore not only form
late, which is a natural consequence of the $\Lambda$CDM cosmology, but the
majority of their mass accretion rates is also faster than an exponential
at low redshifts.

The results presented thus far are for the mean MAH and mean values of
$(\beta, \gamma)$. We find, however, significant dispersions about the mean
behavior that are also important to characterize. For completeness, we show
the distributions of our best-fit $(\beta, \gamma)$ for all halo MAHs in
the Appendix and Fig.~\ref{fig:betagamma}. We also present there an
accurate fitting form that we have obtained for the two-dimensional
probability distribution of $\beta$ and $\gamma$ as a function of halo
mass. This formula can be used to generate a Monte Carlo ensemble of
realistic halo growth histories. The details of the formula, its usage, and
comparison to the Millennium data are described in the Appendix.  We
emphasize that the results presented for the rest of this paper are
obtained from the Millennium haloes directly rather than from this Monte
Carlo realization.

\subsection{MAHs for Haloes at Higher Redshifts}

The MAHs presented thus far are obtained from the main branches of the
descendant haloes at $z=0$. Thus, for a higher redshift $z_1>0$, the
distribution of $M(z_1)$ contains only information about the main branch
progenitors, which is a subset of all the haloes at $z_1$ since many haloes
do not belong to main branches.

Since the formation of higher-redshift galaxies and their host haloes is of
much interest, it is useful to quantify the behavior of MAHs for haloes at
$z_1$, where $z_1>0$. In particular, we ask whether the mean MAH for haloes
of mass $M_1$ at $z_1$ for $z>z_1$ can be related to the MAHs of haloes at
$z=0$ that we have studied thus far.

We find that the mean MAH of haloes of mass $M_1$ at $z_1$ is nearly
identical to the mean MAH of haloes at $z=0$ that satisfy
$M(z_1)=M_1$. That is, the mean MAH for $z>z_1$ of the main branch subset
with mean mass $M(z_1)=M_1$ at $z_1$ is very similar to the mean MAH of the
complete population of haloes with mean mass $M_1$ at $z_1$.  As a specific
example, the mean MAH of the $10^{13} M_\odot$ $z=0$ haloes in the
simulations had the value $M(z=1)=4.5\times10^{12} M_\odot$ at $z=1$. We
find that the mean MAH of these $10^{13} M_\odot$ haloes at
$z>1$ is nearly identical (within 2\%) to the $z>1$ evolution of the mean
MAH of \emph{all} $M_1=4.5\times10^{12} M_\odot$ haloes at $z=1$.  This
property for the mean MAH is in fact a natural consequence of the Markovian
nature of the Extended Press-Schechter theory (see, e.g., Sec 2.3 of
\citealt{White96}).

\begin{figure}
  \centering
    \includegraphics{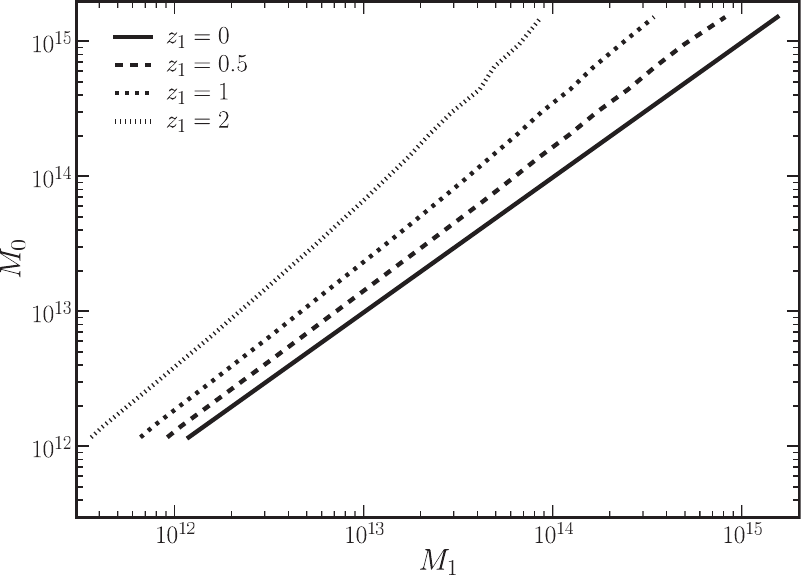}
    \caption{For haloes of mass $M_1$ at redshift $z_1$, the y-axis plots
      their corresponding mean mass $M_0$ today. Four values of $z_1$ are
      shown: 0 (solid), 0.5 (dashed), 1 (dot dashed), and 2 (dotted). This
      mapping allows one to use equation~(\ref{ourfit}) for the MAH of
      higher-redshift haloes (see text). } 
\label{fig:highz_m0}
\end{figure}

This self-similar property implies that in order to study the MAH
properties of haloes with mass $M_1$ at redshift $z_1$, one simply needs to
determine which set of haloes at $z=0$ have $M(z_1)=M_1$. In particular,
one needs to compute the average mass $M_0$ of the haloes at $z=0$ that map
onto $M(z_1)=M_1$ at $z_1$. This mapping is shown in
Fig.~\ref{fig:highz_m0} with $M_1$ along the x-axis and $M_0$ along the
y-axis for $z_1=0,0.5,1,$ and $2$. Note that $M_0=M_1$ at $z_1=0$ by
construction, and as $z_1$ increases, the mass $M_0$ that maps onto some
fixed $M_1$ by redshift $z_1$ also increases.

We note that the mapping in Fig.~\ref{fig:highz_m0} implies that haloes of
some mass $M_1$ at some redshift $z_1>0$ do {\it not} have the same shape
of MAH as haloes of mass $M_0=M_1$ at $z=0$. That is, the MAH of a $10^{13}
M_\odot$ halo at $z=0$ and the MAH of a $10^{13} M_\odot$ halo at $z_1>0$
are not simply related by a shift from $z$ to $z-z_1$ in
equation~(\ref{ourfit}).  This is because haloes at higher $z_1$ have a
{\it relative} formation redshift $z_f-z_1$ that is smaller than haloes of
the same mass at $z=0$. This result is not surprising since haloes of the
same mass at different redshifts in the $\Lambda$CDM model represent
different part of the mass spectrum and are not generally expected to have
identical properties.

We have tested the self-similar property of the fitting form of Z08 (using
their online code) by comparing their mean MAH for $z=0$ $10^{13} M_\odot$
haloes and the MAH for their $M(z=1)$ haloes at $z>1$.  Their latter MAH is
higher than the former by about 15\%, while ours differ by less than 2\%.

\section{Mass Accretion Rates: Mean and Dispersion} \label{marate}

\begin{figure}
  \centering
  \includegraphics{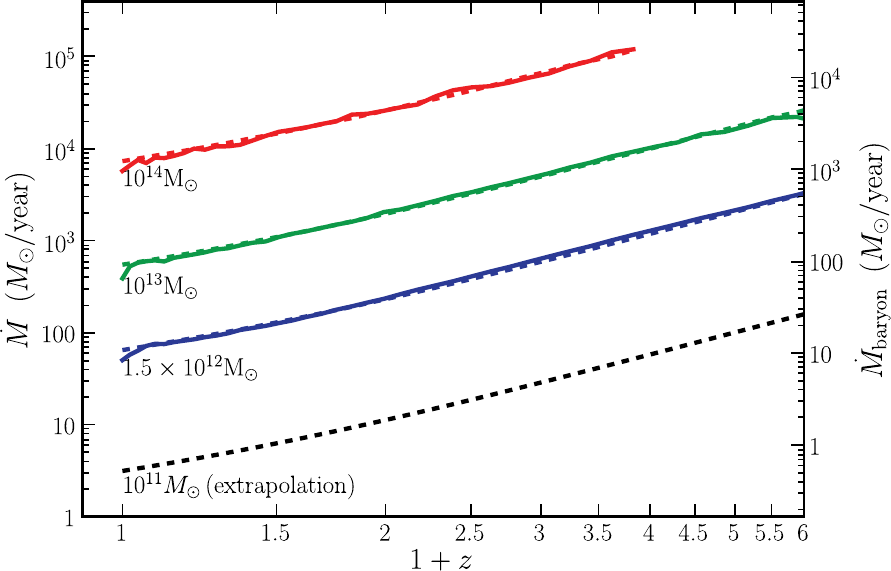}
  \caption{Mean mass accretion rate of dark matter as a function of
    redshift at halo mass $10^{11}, 1.5\times 10^{12}, 10^{13}$ and
    $10^{14} M_\odot$. The solid curves are computed from the Millennium
    haloes (except $10^{11} M_\odot$, which falls below our resolution
    limit of 1000 particles per halo) at a given mass ($\pm 20\%$ range);
    the dashed curves show the accurate fit provided by
    eq.~(\ref{Mdotfit}). The right side of the vertical axis labels the
    mean accretion rate of baryons, $\dot{M}_b$, assuming a cosmic
    baryon-to-dark matter ratio of $\sim 1/6$.  The slight dip in $\dot{M}$
    at $z=0$ is due to the artificial edge effect inherent in the stitch-3
    algorithm used to process the FOF merger trees.}
\label{fig:meanrate}
\end{figure}

\begin{figure*}
  \centering
  \includegraphics{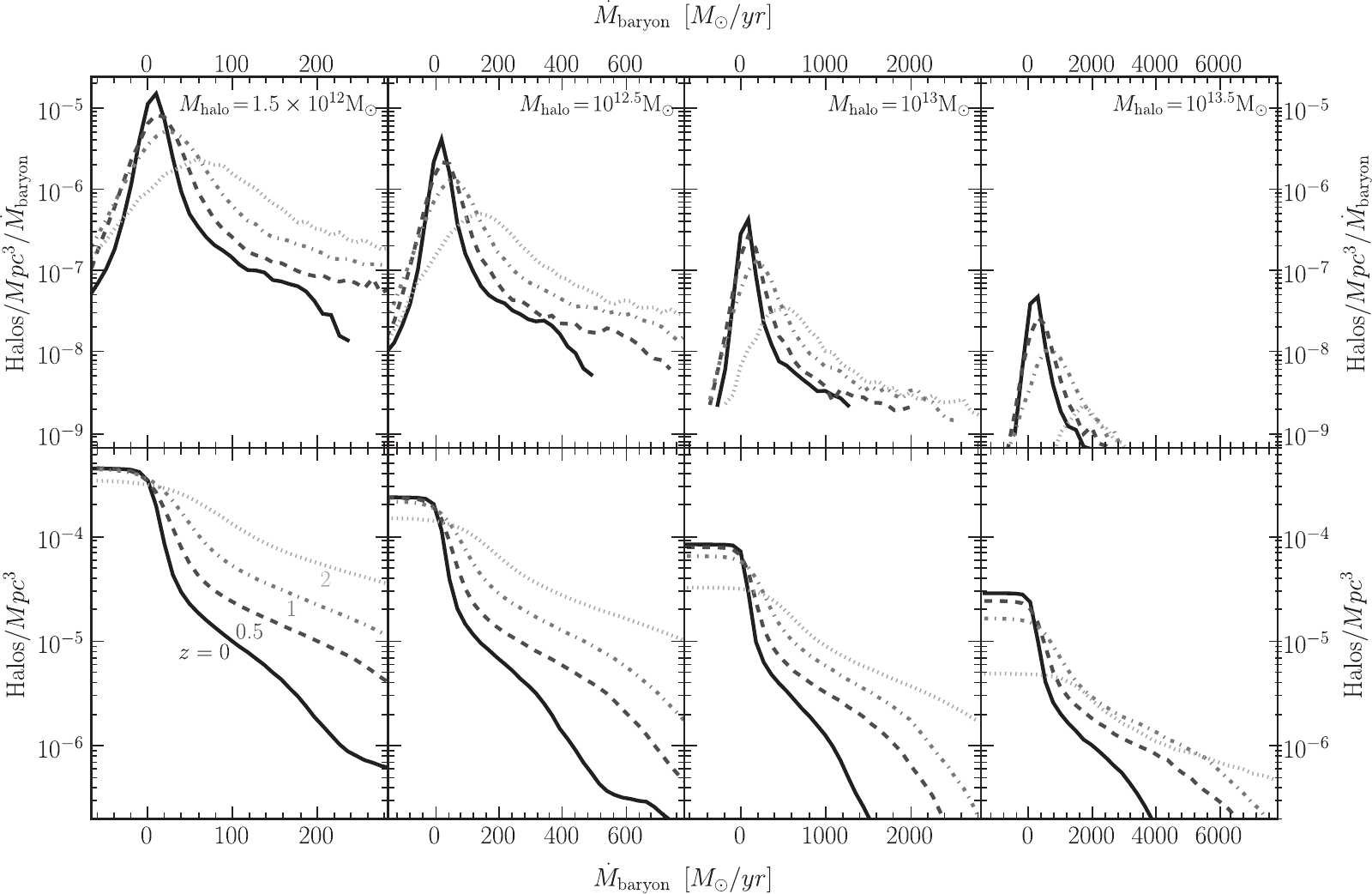}
  \caption{Differential (top) and cumulative (bottom) distributions of the
    accretion rates of cosmic baryons, $\dot{M}_b$, for four halo masses
    (left to right).  Within each panel, the accretion rates at four
    redshifts $z=0$ (solid), 0.5 (dashed), 1 (dashed dotted), and 2
    (dotted) are shown, where the distributions are seen to broaden
    significantly with increasing $z$. The vertical axis labels the number
    of haloes per comoving Mpc$^3$ at or above a given $\dot{M}_b$.}
\label{fig:rate}
\end{figure*}

Having quantified $M(z)$ in Sec.~3, we now examine its time derivative --
the mass accretion rate -- in more detail. In particular, we would like to
obtain a general formula for the mean accretion rates of dark matter for a
wide range of halo mass and redshift. To achieve this, we note that our
analytical form in equation~(\ref{ourfit}) for individual halo MAHs gives:
\begin{equation}	
	\frac{\dot{M}}{M} = 0.10 h \, {\rm Gyr}^{-1} 
     \left[ \gamma (1+z) - \beta \right] \sqrt{\Omega_m (1+z)^3 + \Omega_\Lambda} 
\label{MdotM} 
\end{equation}
where $\Omega_m$ and $\Omega_\Lambda$ are the present-day density
parameters in matter and the cosmological constant, and we have assumed
$\Omega_m + \Omega_\Lambda=1$ (used in the Millennium simulation) and
matter-dominated era in computing $dz/dt$. As shown in Sec.~3, the
parameters $\beta$ and $\gamma$ in equation~(\ref{MdotM}) generally depend
on the halo mass. We find, however, that the mass dependence follows a
simple power law independent of the redshift, and the simple analytic form
in equation~(\ref{MdotM}) provides an excellent approximation for the {\it
  mean} mass accretion rate as a function of redshift and halo mass:
\begin{eqnarray}
	\left< \dot{M} \right> &=& 42 \, M_\odot {\rm yr}^{-1} 
         \left( \frac{M}{10^{12} M_\odot} \right)^{1.127} \nonumber\\
	&& \times (1 + 1.17 z) \sqrt{\Omega_m (1+z)^3 + \Omega_\Lambda} \,. 
\label{Mdotfit} 
\end{eqnarray}
For completeness, the best fit for the \emph{median} growth rate computed in the Millennium simulation is
\begin{eqnarray}
	\left< \dot{M} \right>_{\rm median} &=& 24.1 \, M_\odot {\rm yr}^{-1} 
         \left( \frac{M}{10^{12} M_\odot} \right)^{1.094} \nonumber\\
	&& \times (1 + 1.75 z) \sqrt{\Omega_m (1+z)^3 + \Omega_\Lambda} \,. 
\label{MdotfitMedian} 
\end{eqnarray}
We note that the overall amplitude of the mean is higher than the median due to the long, positive, $\dot{M}$ tail (see Fig. \ref{fig:rate}).

Fig.~\ref{fig:meanrate} compares the mean accretion rates of dark matter in
$M_\odot$ per year computed from the Millennium simulation (solid curves)
and this formula (dashed curves) for haloes of mass $10^{12} M_\odot$ to
$10^{15} M_\odot$ over the redshift range of 0 and 5.  The overall trend of
the accretion rate is such that $\dot{M}/M$ has a weak dependence on $M$
($\propto M^{0.127}$), and its dependence on redshift is approximately
$(1+z)^{1.5}$ at low $z$ and $(1+z)^{2.5}$ at $z > 1$. This $z$-dependence
is motivated by our 2-parameter form for $M(z)$ and is more accurate than
the simple power law used in \citet{Genel08}, \citet{Neistein06}, and
\citet{Neistein08}; our $z\sim0$ value, on the other hand, is consistent
with theirs to within 20\%.  We have also computed $\dot{M}$ from the
fitting form for the median MAH in the recent preprint by Z08.  We found
their $\dot{M}$ to have a slightly steeper $z$-dependence than our
equation~\ref{MdotfitMedian} where their median
value is within 20\% of our median $\dot{M}$ at $z\sim 0$ but exceeds ours
by a factor of $\sim2$ at $z\sim4$.

Along the right side of the vertical axis of Fig.~\ref{fig:meanrate}, we
label the corresponding mean accretion rates of baryons, $\dot{M}_b$,
assuming a cosmic baryon-to-dark matter ratio of $\Omega_b/\Omega_m \approx
1/6$. The results shown should be a reasonable approximation for the mean
rate of baryon mass that is being accreted at the virial radius of a dark
matter halo of a given mass. Fig.~\ref{fig:meanrate} and
equation~(\ref{Mdotfit}) indicate that this rate is $\dot{M}_b\approx 7 \,
M_\odot {\rm yr}^{-1}$ for $10^{12} M_\odot$ haloes today, and it increases
to 27, 69, and 140 $M_\odot {\rm yr}^{-1}$ for $10^{12} M_\odot$ haloes at
$z=1$, 2, and 3, respectively. Since the infalling baryons are a reservoir
for the gas that fuels star formation, it is interesting to compare
$\dot{M}_b$ with the mean star formation rates of different types of
galaxies, e.g., $\dot{M}_*\sim 4 M_\odot {\rm yr}^{-1}$ for the Milky Way
(e.g., \citealt{Diehl06}), suggesting that about half of the infalling
$\dot{M}_b\approx 7 \, M_\odot {\rm yr}^{-1}$ for Galactic-size haloes
needs to be converted into stars.  The relations among these different
accretion rates and the implications will be investigated in a subsequent
work.

Having determined the mean rates, we show their distributions and
dispersions in Fig.~\ref{fig:rate}.  Four redshifts, $z=0$, 0.5, 1, and 2,
and four ranges of halo masses (left to right panel) are shown. Both the
differential (top panels) and cumulative (bottom panels) distributions of
$\dot{M}_b$ are plotted for comparison.  Within each panel, the
distribution of $\dot{M}_b$ at a given halo mass is seen to broaden
significantly with increasing redshift.  For instance, the (comoving)
number density of $1.5\times 10^{12} M_\odot$ haloes with $\dot{M}_b > 250
M_\odot$ yr$^{-1}$ increases dramatically from $5\times 10^{-7}$ Mpc$^{-3}$ at
$z=0$ to $5\times 10^{-5}$ Mpc$^{-3}$ at $z=2$.  At a given redshift, the
distribution of $\dot{M}_b$ also broadens with increasing halo mass,
although the distribution (and dispersion) of the ratio $\dot{M}_b/M_b$ is
largely independent of mass. The latter is similar to the weak mass
dependence of the mean $\dot{M}/M$ given by equation~(\ref{Mdotfit}).

\section{Formation Redshifts: Mean and Dispersion} 
\label{formation_redshift}

It is well established that on average, more massive haloes form later than
less massive haloes in the $\Lambda$CDM cosmology. The Millennium database
provides sufficient statistics for us to quantify the distributions of the
formation redshift $z_f$ and its mean {\it and} scatter over a wide range
of halo masses ($\sim 10^{12}$ to $\sim 10^{15} M_\odot$). The formation
redshift, along with the late-time growth rate $\beta-\gamma$, can be
thought of as two physically motivated quantities parameterizing the halo
MAH.
\begin{figure*}
  \centering
  \includegraphics{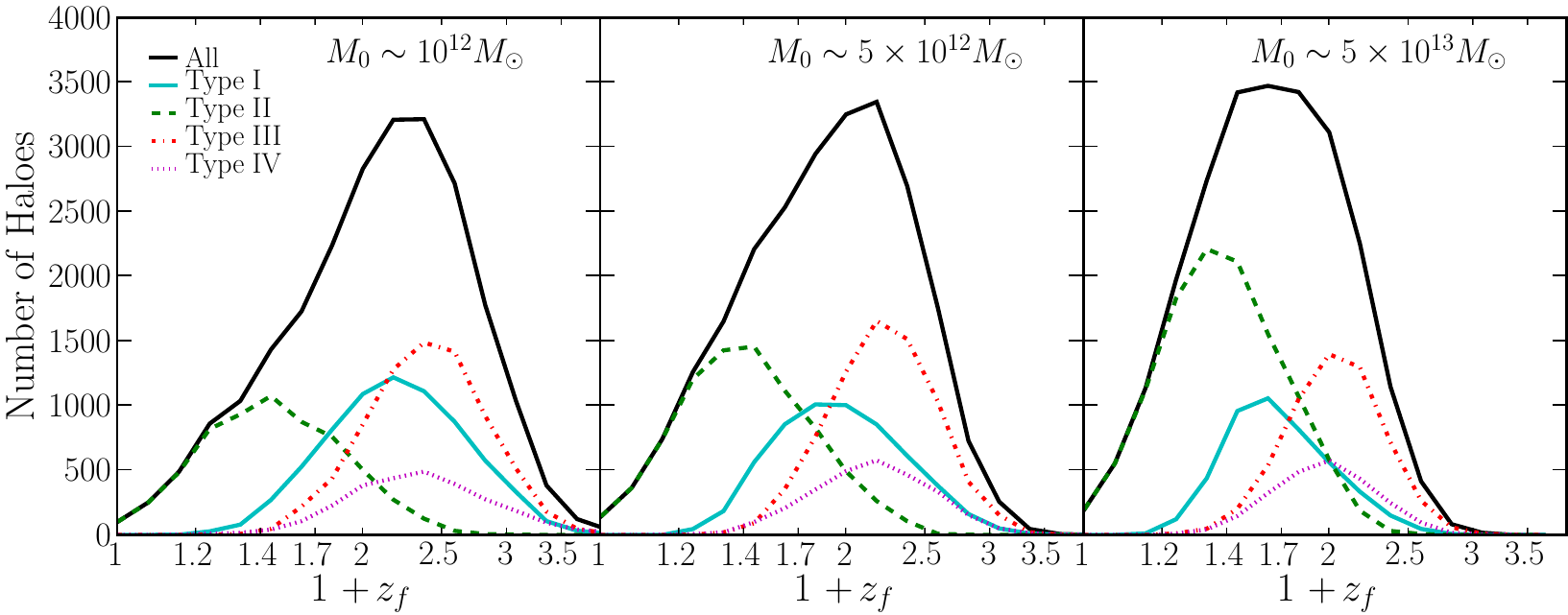}
  \caption{Distributions of the formation redshift $z_f$ for three mass
    bins (left to right). Within each mass bin, the $z_f$ distribution for
    all haloes is plotted (solid black), as well as the distribution for
    each halo type. While the relative amplitudes of the distributions do
    change from one mass bin to another, the overall shapes remain similar
    across all masses.} \label{fig:zf_panels}
\end{figure*}

\begin{figure*}
  \includegraphics{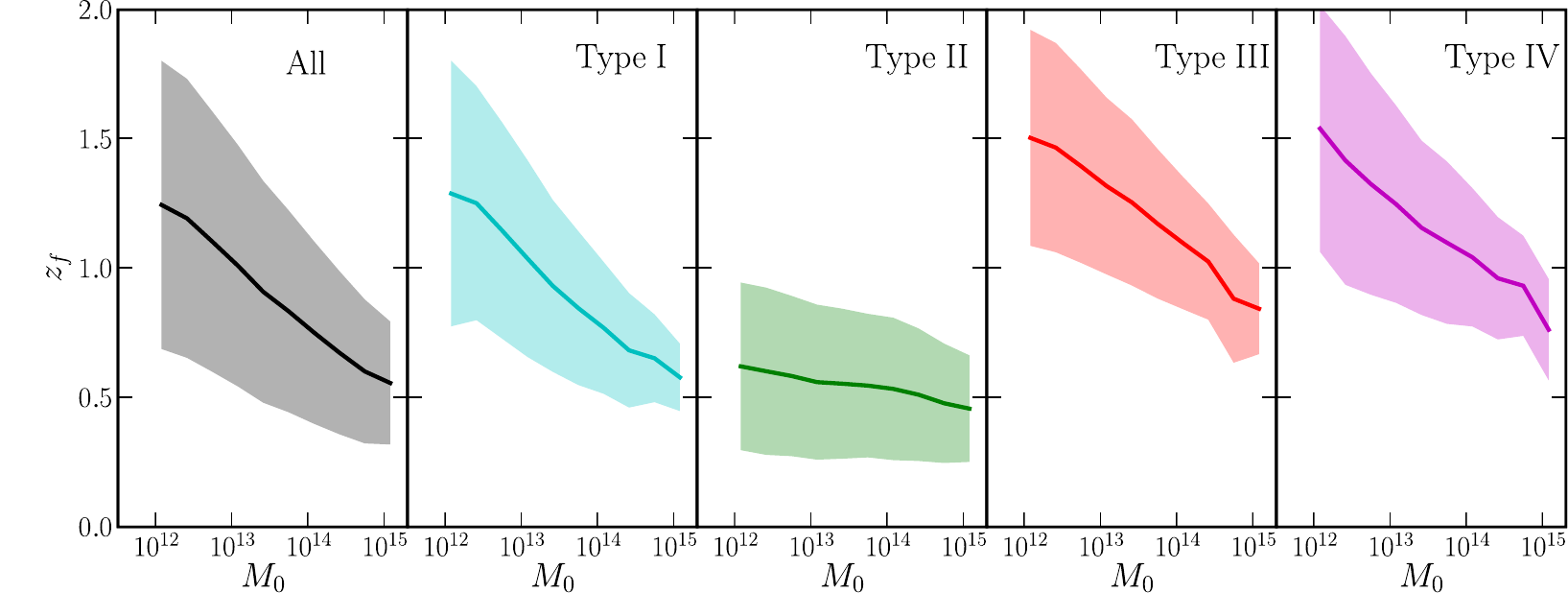}
  \caption{Mass dependence of the mean (solid curves) and one standard
    deviation scatter (shaded regions) of the formation redshifts of the
    Millennium haloes. More massive haloes on average form more recently,
    but the scatter is large. One exception is Type II haloes that have a
    mean $z_f$ of $\approx 0.5$ independent of mass.  Fits to the mean and
    scatter of $z_f$ as a function of mass are given in
    Table~\ref{table:zf_b-a_fits}.  }
\label{fig:m0_zf_panels}
\end{figure*}
\begin{table*}
	\begin{tabular}
		{l c c c c} & \multicolumn{2}{c}{$z_f$ vs. $M_0$} & \multicolumn{2}{c}{$\gamma - \beta$ vs. $M_0$} \\
		\hline \hline & $\left< z_f \right>$ & $\sigma_{z_f}$ & $\left< \gamma-\beta \right>$ & $\sigma_{(\gamma-\beta)}$ \\
		\hline Overall & $-0.24x + 1.26$ & $-0.11x + 0.58$ & $-0.25x -0.29$ & $0.14x + 0.64$\\
		Type I & $-0.25x + 1.30$ & $-0.12x + 0.51$ & $-0.22x-0.45$ & $0.03x + 0.21$ \\
		Type II & $-0.05x + 0.62$ & $-0.04x + 0.34$ & $-0.12x -1.15$ & $0.14x + 0.60$\\
		Type III & $-0.23x + 1.56$ & $-0.08x + 0.43$ & $-0.03x -0.20$ & $0.01x + 0.11$\\
		Type IV & $-0.23x + 1.56$ & $-0.08x + 0.43$ & $0.03x + 0.15$ & 0.36 \\
	\end{tabular}
	\caption{ Linear fits for the mass dependence of the mean formation redshift $z_f$, mean $\gamma - \beta$, and their respective 1$\sigma$ scatter about the mean, where $x \equiv \log_{10}(M_0 / 10^{12} M_{\sun})$. } \label{table:zf_b-a_fits} 
\end{table*}

The distributions of $z_f$ for each type of MAHs for three halo mass bins
are plotted in Fig.~\ref{fig:zf_panels}. The overall trend of decreasing
$z_f$ with increasing halo mass is evident across the three panels. Within
each panel, a correlation between $z_f$ and the MAH type is clearly
seen. Nearly all haloes in the smallest few $z_f$ bins are Type II. This
means that despite the fact that Type II dominates the highest mass bins,
the haloes that constitute Type II are not merely the especially massive
haloes which formed late, but also include less massive haloes which formed
late. On average, a Type II halo has a formation redshift 0.5 smaller than
a typical halo. To a lesser degree, Type III and Type IV are also distinct
from the overall distribution. Both tend to form early, Type III more so
than Type IV, and together the two types account for most of the haloes
that formed early.

Fig.~\ref{fig:m0_zf_panels} shows the mean formation redshift $z_f$ as a
function of halo mass for all 478,781 $z=0$ Millennium haloes (leftmost
panel) as well as for each type of MAH. As the scatter about the line is,
to a good approximation, Gaussian, the 1$\sigma$ range about the line is
also provided in the plots in each panel (light shaded areas). From these
shaded areas, it is clear that there is considerable scatter for the
overall distribution. The relationship between $M_0$ and $z_f$ is different
from the overall distribution for all types except Type I, which suggests
that the types discriminate by formation redshift to some extent. Also note
that the separation of haloes into types also produces more limited scatter
about the mean.

To approximate the mass dependence of the mean and scatter of $z_f$, we use
the linear form
\begin{equation}
	\left< z_f \right> = a \log_{10} \frac{M_0}{10^{12}M_\odot} + b \,,\quad 
        \sigma_{z_f} = c \log_{10} \frac{M_0}{10^{12}M_\odot} + d 
\label{zfit}
\end{equation}
and find it to fit the simulation data
accurately. Table~\ref{table:zf_b-a_fits} lists the best-fit coefficients
for all the halo MAHs (above 1000 particles at $z=0$) and for each of the
four types of MAHs shown in Fig.~\ref{fig:m0_zf_panels}.
Table~\ref{table:zf_b-a_fits} also includes the same fit performed for the
fit parameters ($\beta - \gamma$). The mean formation redshifts differ
significantly among the types, with $\left< z_f \right> \approx 0.6, 1.3$,
and 1.5 for Type II, I, and III (plus IV), respectively, for galaxy-size
haloes. The dependence of $\left< z_f \right>$ on mass is noticeably weak
for Type II; the other types show similar mass dependence, where $d\left<
  z_f \right>/d\log M$ ranges from $-0.23$ to $-0.25$.

With the relationships between formation redshift and mass for each type,
we can look at how these dependences relate to the basic halo
characteristics given in Table~\ref{table:types}. Recall that Type II
haloes were marked by steep growth at late times, which is captured by the
very negative value of $\beta - \gamma$. Type III haloes, on the other
hand, have small values for $\beta - \gamma$, and thus grow slowly at late
times. The relationships shown in Fig.~\ref{fig:m0_zf_panels} are then no
surprise. Type II haloes are also associated with late formation times,
while Type III haloes tend to have formed quite early.

\section{Correlations with Halo Environments and Major Merger Frequencies}

Thus far we have discussed how the halo MAHs and mass accretion rates vary
with halo mass and redshift. We have also shown that the mean $z_f$ depends
on halo mass strongly, but the scatter in $z_f$ does not depend strongly on
the MAH type nor halo mass. In this section, we investigate if the mean and
scatter in $z_f$ are correlated with quantities other than halo mass. In
particular, we ask if the shapes of MAHs (1) differ systematically between
underdense vs overdense regions, and (2) are correlated with the time and
frequency of major mergers and mass brought in by these events during a
halo's lifetime.

\subsection{Environment}

An extensive discussion and tests of halo environments can be found in
\citet{FM09}. Four definitions of a halo's local environment based on the
local mass density centered at the halo were compared. Three of them were
computed using the dark matter particles in a sphere of radius $R$ centered
at a halo, either with or without the central region carved out; the fourth
definition was computed using the masses of only the haloes rather than all
the dark matter. Here we use $\dRfof$, computed by subtracting out the FOF
mass $M$ of the central halo within a sphere of radius $R$:
\begin{equation}
	\dRfof \equiv \delta_R - \frac{M}{V_R \bar{\rho}_m} \,, \label{delta3} 
\end{equation}
where $V_R$ is the volume of a sphere of radius $R$, and $\delta_R$ is the
mean mass overdensity within $R$.  This measure makes no assumption about the
central halo's shape. By taking out the mass of the central halo itself,
this density was shown to be a more robust measure of the environment {\it
  outside} of a halo's virial radius. Otherwise, the halo mass itself
dominates the density centered at massive haloes (see Fig.~1 of
\citealt{FM09}), and it becomes difficult to distinguish whether any
correlations are due to the mass or the larger-scale environment in which
the halo resides.

\begin{figure}
  \centering
  \includegraphics{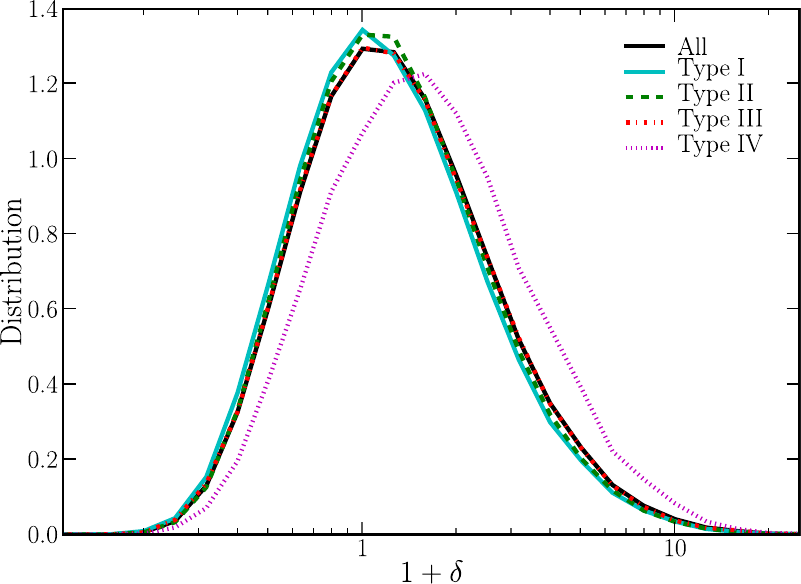}
  \caption{Normalized distributions in $\delta$ are plotted for all haloes,
    as well as each type of halo. The most prominent feature is the clear
    separation between Type IV and all other types. The center of the Type
    IV is well to the right of the other types, meaning that Type IV haloes
    are predominantly found in denser environments than any other
    type.} 
\label{fig:delta_dist}
\end{figure}

Fig.~\ref{fig:delta_dist} shows the distribution of the environmental
densities (evaluated at $z=0$) for haloes within each MAH type and the
total population. The distribution for Type IV haloes is quite distinct
from all other distributions and is offset towards higher densities. Since
Type IV haloes experience very little mass growth or even mass loss at late
times, denser environments appear to impede mass accretion onto haloes.

\subsection{Mass Growth due to Major Mergers}

Major mergers are more rare than minor mergers, but they can contribute to
a significant fraction of a halo's final mass, and have a strong impact on
halo structures and galaxy properties such as the star formation rate. 

\begin{figure}
  \centering
  \includegraphics{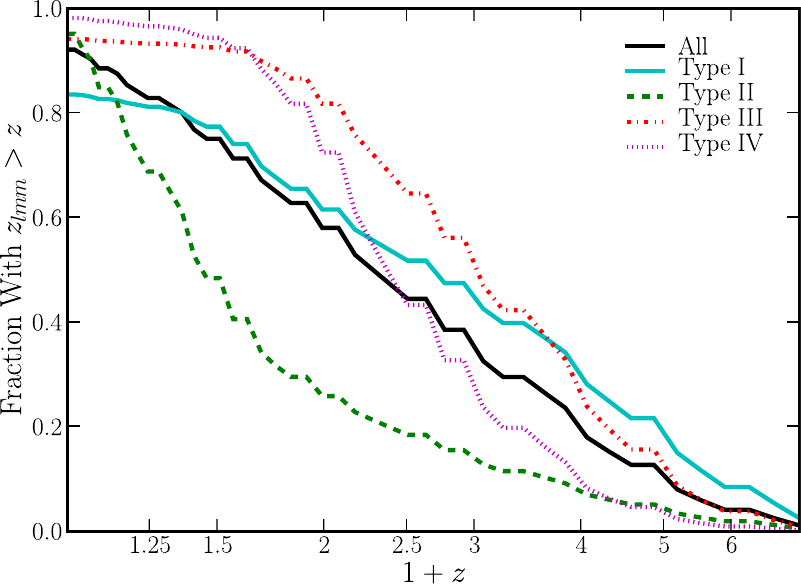}
  \caption{Distribution of the redshift of the last merger major for the
    four types of haloes (colored curves) and all $z=0$ haloes (solid
    black).  The majority of the $z=0$ haloes with Type II MAH (dashed
    green curve) have experienced a major merger (of mass ratio $>0.33$)
    very recently, whereas the last major merger occurred earlier than
    $z=1$ for more than 75\% of Type III and IV haloes.}
\label{fig:lmm}
\end{figure}

A useful quantity for assessing the role of major mergers on halo
MAHs is $z_{lmm}$, the redshift of the last major merger in a halo's
history. Fig.~\ref{fig:lmm} shows the fraction of haloes whose last major
merger occurred at or before $z$. Type II haloes are seen to experience a
major merger in the much more recent past than the other types: about 65\%
of them had encountered a major merger within redshifts 0 and 0.3, and only
25\% of them had their last major merger before $z=1$. In sharp contrast,
only about 5\% of Type III and IV haloes had a major merger later than
$z=0.3$, and over 75\% of them had their last major merger earlier than
$z=1$.

\begin{figure}
  \centering
  \includegraphics{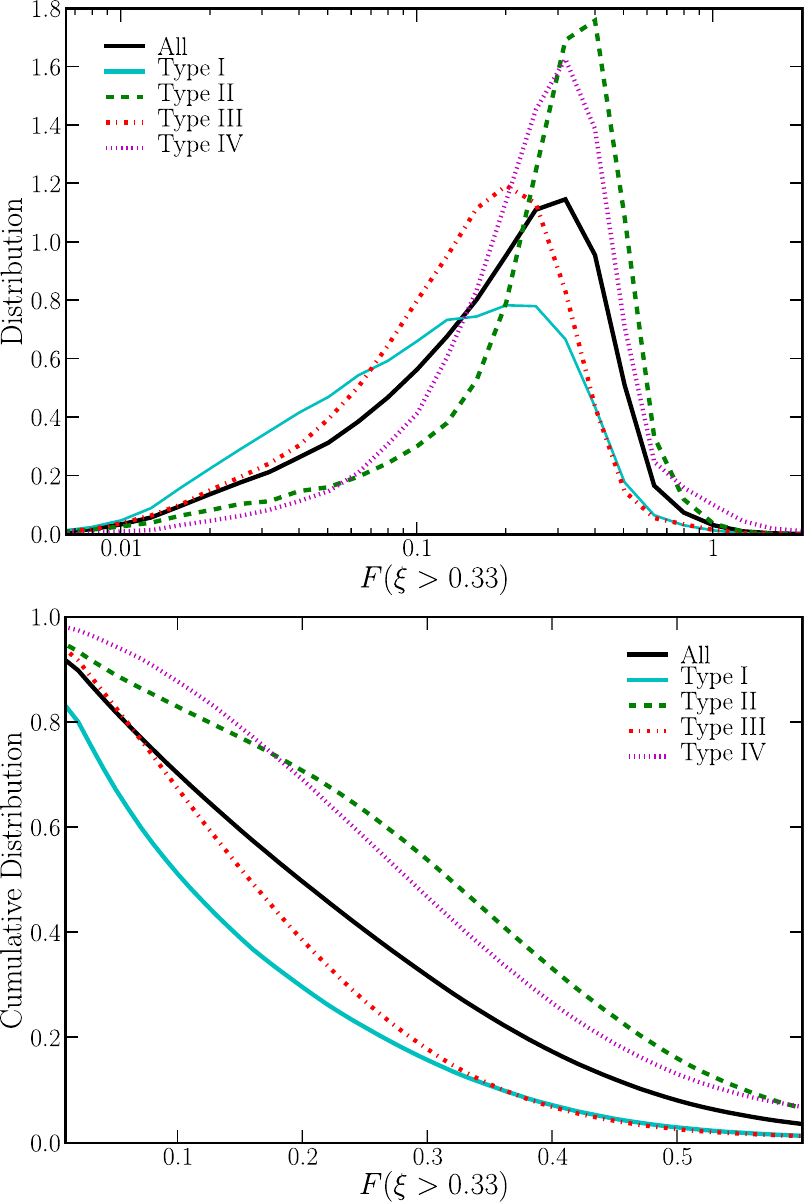}
  \caption{Differential (top) and cumulative (bottom) distributions of the
    fraction of halo mass gained from major mergers of mass ratio $\xi >
    0.33$. Much like the $z_f$ distributions, each Type is quite distinct
    from the overall distribution. Type I and III haloes feature few major
    mergers, while Type II and Type IV haloes tend to be dominated by major
    mergers. Type IV mergers also feature a noticeable rise in the highest
    bin, due to having gained more mass via major mergers across their
    history than they currently have.} \label{fig:f_dists}
\end{figure}

Another useful parameter for quantifying the role of major mergers in its MAH 
is $F(\xi > \ximin)$, which is the fraction of mass at $z_0$ that
came from mergers above some progenitor mass ratio $\ximin$. We choose to
define the mass ratio in relation to the mass of the progenitor at the time
of merger, as opposed to being defined in relation to the halo's present
mass.  The exact value of $F$ is strongly dependent upon the choice of
$\ximin$. The overall features, however, do not change significantly, so
different values for $\ximin$ only change the values for $F(\xi > \ximin)$,
but leave the overall characteristics in place.

Fig.~\ref{fig:f_dists} shows the differential (top) and cumulative (bottom)
distribution of the major merger mass fraction, $F(\xi > 0.33)$ for each
type of MAHs. Type II haloes (dashed curves), which feature steep growth at
late times, are seen to have the highest $F$ among the four types. The
distribution peaks at $F \approx 0.4$, indicating that $\sim 40$\% of their
final mass was acquired through major mergers. Type I haloes (solid cyan
curves), by contrast, feature a dearth of major mergers, which is
unsurprising given the fact that large mergers are poorly handled by the
simple exponential. Likewise, Type IV haloes (dotted curves), which grow
quickly early on only to lose mass at late times, are dominated by major
mergers with a number of Type IV haloes even having $F(\xi > \ximin) > 1$.
This occurs when a halo has less mass presently than it has gained overall
via major merger. Type III haloes (dotted dashed curves) tend to be
relatively major merger free, which is again to be expected due to the
decelerating growth of Type III haloes at later times during which much of
the overall mass is accreted.

\section{Conclusions} \label{conclusion}

We have examined the mass growth histories of $\sim 500,000$ $z=0$ haloes
and their progenitors in the Millennium simulation.  The two-parameter
function in equation~(\ref{ourfit}) provides a reasonable fit for the MAHs
of these haloes, as shown in Fig.~\ref{fig:mass_sum2}.  The mean mass
accretion rate of dark matter (and baryons) as a function of halo mass and
redshift is well approximated by equation~(\ref{Mdotfit}), as shown in
Fig.~\ref{fig:meanrate}.  The distributions of $\dot{M}$ are broad, and the
number density of high-$\dot{M}$ haloes increases sharply with increasing
$z$ at a given halo mass (see Fig.~\ref{fig:rate}).  
The mean halo formation redshift as a function of
mass is given by equation~(\ref{zfit}) and Fig.~\ref{fig:m0_zf_panels}.

To facilitate the analysis of the halo MAH, we have classified $M(z)$ into
four types based on their shapes.  We have shown that only 20 to 30\% of
the Millennium haloes follow an exponential form (``Type I'') in their mass
accretion history $M(z)$. Only one parameter is needed to specify their
MAH, e.g., the formation redshift $z_f$. The formation redshift depends
strongly on halo mass, as expected for hierarchical cosmological models
such as the $\Lambda$CDM. The median $z_f$ ranges from 1.3 for $10^{12}
M_\odot$ haloes to 0.6 for $10^{15} M_\odot$ haloes, and is dispersed over
a range roughly equal to the median value for all masses.

About 20\% of galaxy-size and 60\% of cluster-size haloes have late-time
growth that is steeper than an exponential (``Type II''). These haloes are
formed more recently, with a median $z_f$ of about 0.5 for all masses. The
redshift at which they experience the last major merger is also
significantly later than the exponential haloes: about 50\% of them have
had the last major merger between $z=0$ and 0.3, as opposed to 10\% of the
rest of the haloes, including exponential haloes. Correspondingly, a higher
fraction of Type II haloes' final mass is acquired through major mergers,
e.g. 60\% of these haloes obtained more than 30\% of their final mass from
major mergers, whereas a little over 30\% of all haloes obtained more than
30\% of their final mass from major mergers, and fewer than 20\% of
exponential haloes did.

The rest of the haloes have stunted late-time growth relative to an
exponential form. The median $z_f$ ranges from 1.5 at low mass to 0.8 at
high mass.  These haloes can be further separated into two groups (Type III
and IV), where the two are primarily distinguished by the roles that major
mergers have played in their growth; that is, Type III haloes tend to
experience few major mergers, whereas Type IV haloes grew predominantly
from major mergers at early redshifts. The MAHs of the two can be
distinguished by the sharpness in the downturn of late time growth. Type IV
haloes also live in somewhat denser environments, where the stronger tidal
fields and more frequent interactions may have contributed to rapid
accretion at early times followed by a slow down of their late time mass
growth.

Despite this diverse behavior of halo MAHs, we have found the individual
$M(z)$ to be well fit when a second parameter is introduced
(eq.~\ref{ourfit}).  To quantify the statistics of $M(z)$, we have provided
fits to the joint probability distribution of the two MAH parameters
$\beta$ and $\gamma$ in the Appendix. These can be used to generate
realizations of halo mass growth histories in semi-analytic models of
galaxy formation that incorporate realistic scatters about the mean trends.

\section*{Acknowledgments}

We thank Simon White and Phil Hopkins for useful comments. The Millennium
Simulation databases used in this paper and the web application providing
online access to them were constructed as part of the activities of the
German Astrophysical Virtual Observatory.

\bibliographystyle{mn2e} 
\bibliography{MFM09}

\begin{appendix}
\section{Joint Distribution of $\beta$ and $\gamma$}

\begin{figure*}
  \centering
  \includegraphics{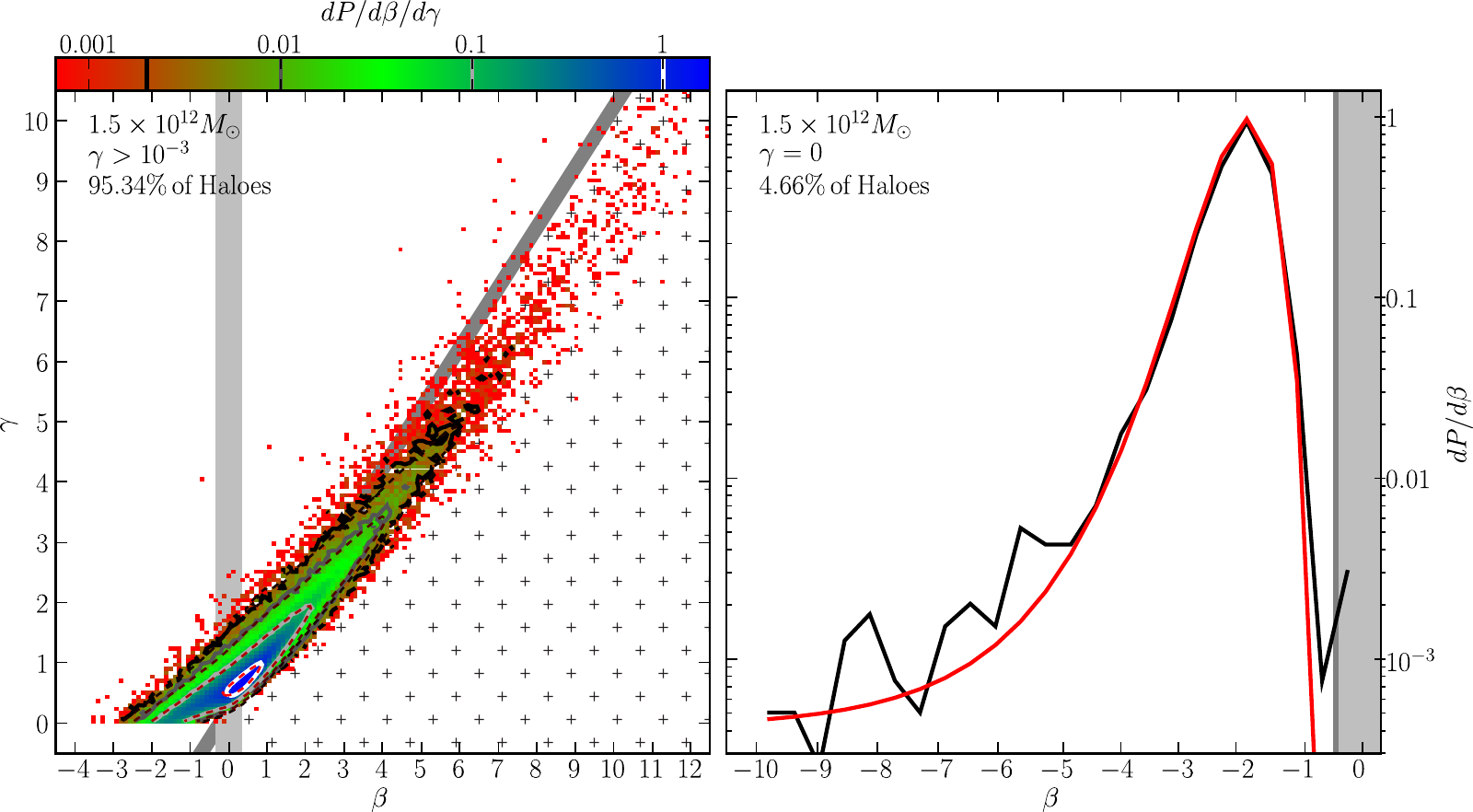}
  \caption{Probability distributions of $\beta$ and $\gamma$ for haloes
    of mass $1.5\times 10^{12} M_\odot$. The left panel is the PDF of
    $\beta$ and $\gamma$ for haloes with $\gamma>10^{-3}$ (95.34\% of all
    haloes). The shaded 2D histogram presents the distribution of $\beta$
    and $\gamma$ obtained from Millennium (see color scale for units) with
    corresponding contours drawn at the 0.005, 0.01,0.1 and 1 levels (black
    to white). The appropriately normalized fitting form $D_{2D}$ is
    overlaid as red dashed contour lines. The right panel is the PDF of
    $\beta$ for haloes with $\gamma=0$ (4.66\%). The appropriately
    normalized fitting form $D_{1D}$ is overlaid in red. For both panels,
    the background shading corresponds to regions of phase space denote
    type I (grey), type II (red), type III (green) and type IV (blue)
    MAHs.} \label{fig:betagamma}
\end{figure*}

\begin{figure*}
  \centering
  \includegraphics{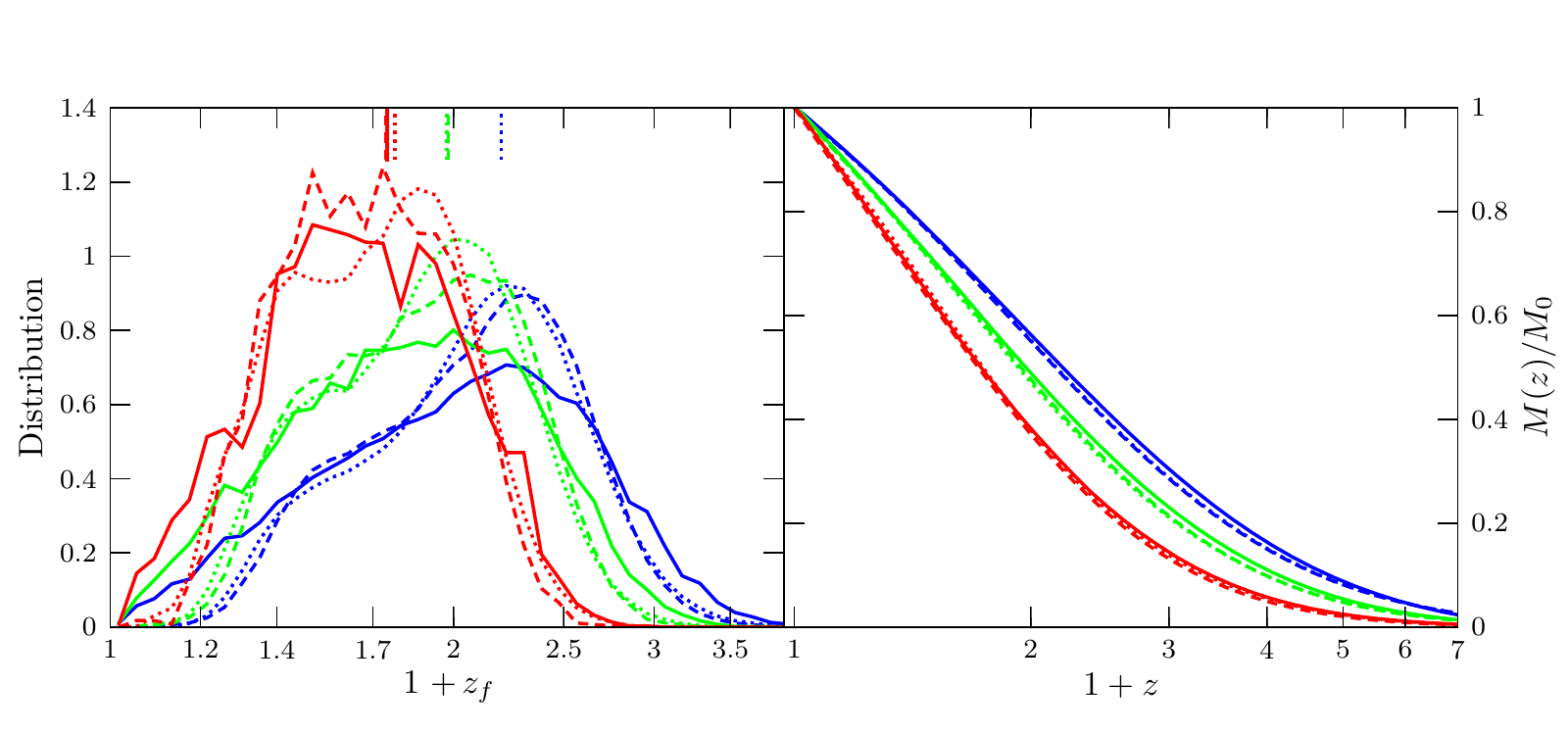}
  \caption{ Distribution of the formation redshifts (left) and the mean
    MAHs (right) from the simulation directly (solid), the ($\beta,\gamma$)
    fits to the simulation MAHs (dashed), and a Monte Carlo ensemble of
    300,000 halos per mass bin generated using equations~(A1)-(A3)
    (dotted).  In each panel, three mass bins are shown:$10^{12}M_\odot$
    (blue), $10^{13}M_\odot$ (green), and $10^{14} M_\odot$ (red).  Note
    that the dashed and dotted curves are almost indistinguishable in the
    right panel.}
\label{fig:MC}
\end{figure*}

We have seen that halo MAHs are well-fit by equation~(\ref{ourfit}) with
two parameters $\beta$ and $\gamma$. Applying this fit to haloes in the
Millennium simulation yields a joint distribution of $\beta$ and
$\gamma$. In this appendix we provide a fitting form to this distribution
as a function of $\beta$, $\gamma$, and halo mass that is intended to allow
the reader to generate rapidly a mock catalog of MAH tracks. We find that a
straightforward rejection method can generate 300,000 mock MAH trajectories
in under a minute on a standard laptop. The mean properties of the
resulting trajectories match the mean properties of the Millennium
trajectories at the 10\% level. The fitting forms presented below are
chosen for the practical purpose of matching the underlying distribution as
closely as possible.
	
We find that 95.34\% of the haloes occupy a smooth region in the ($\beta,
\gamma$) plane shown in the left panel of Fig.~\ref{fig:betagamma}. The
remaining 4.66\% of the haloes live along a distinct line with $\gamma=0$
and $\beta\leq0$, where the distribution of $\beta$ is shown in the right
panel of Fig.~\ref{fig:betagamma}. That is, their MAHs are better
approximated by a power law in $1+z$ rather than an
exponential. Interestingly, this 95.34\% vs 4.66\% division is independent
of halo mass, even though the shape of the distributions generally depends
on mass. For accuracy, we choose to separate the distribution of $\beta$
and $\gamma$ into these two components and fit to them separately.
	
For the 4.66\% of haloes with $\gamma=0$, their $\beta$ distribution is
well approximated by
\begin{equation}
  \frac{dP}{d\beta} \propto e^{-X^2} 
\end{equation}
where
\begin{equation}
  X=7.443 \, e^{0.6335\beta+0.2626M^{0.1992}} - 2.852M^{-0.05412} 
\end{equation}
and $M\equiv M_{\rm halo}/10^{12} M_\odot$. This fit is valid in the range
$-10<\beta\leq0$.
	
For the 95.34\% of haloes with $\gamma > 10^{-3}$, the joint distribution
in $\beta$ and $\gamma$ is well approximated by
\begin{equation}
  \frac{dP}{d\beta d\gamma} \propto e^{-(X M^{-0.05569})^2-(Y M^{-0.05697})^2} 
\end{equation}
where
\begin{eqnarray*}
		X&=&(-1.722-0.1568\beta+0.007592\beta^2)(1-T_2)+\\
		&&(1.242+0.3138\beta-0.01399\beta^2)T_2\\
		Y&=&13.39 [1\!-\!1.224\tanh(1.043Y')] (1\!-\!0.08018\beta) \\
		Y'&=&\gamma-(28.85+0.4537\beta)(1-T_1)-\\
		&&(28.38+0.7624\beta)T_1+29.21M^{-0.001933}\\
		T_1&=& 0.5 [1+\tanh(1.174\beta)]\\
		T_2&=& 0.5 [1+\tanh(0.7671\beta-0.1269)] \,. 
\end{eqnarray*}
This is valid in the range $-8<\beta<12$, $0<Y'<3$. Since the rejection
method does not require a normalized PDF for input, we leave these
probability distributions unnormalized.
	
Fig.~\ref{fig:MC} illustrates that Monte Carlo realizations generated from
the probability distributions above (dotted curves) reproduce accurately
the formation redshift distributions (left panel) and the mean MAHs (right
panel) obtained from the $(\beta,\gamma)$ fits to the Millennium MAHs
(dashed curves), and both match closely the results computed directly from
the Millennium simulation (solid curves).
	
\end{appendix}

\label{lastpage}

\end{document}